\documentclass[conference]{IEEEtran}
\IEEEoverridecommandlockouts
\usepackage{cite}
\usepackage{amsmath,amssymb,amsfonts}
\usepackage{algorithmic}
\usepackage{graphicx}
\usepackage{textcomp}
\usepackage{xcolor}
\usepackage{multirow} 
\usepackage{multicol} 
\usepackage{caption}
\usepackage{float}
\usepackage{arydshln}
\usepackage{subfigure}
\usepackage{enumerate}
\usepackage{marvosym}
\usepackage{colortbl}
\usepackage{url}

\def\BibTeX{{\rm B\kern-.05em{\sc i\kern-.025em b}\kern-.08em
    T\kern-.1667em\lower.7ex\hbox{E}\kern-.125emX}}
\begin{document}

\title{
AGMI: Attention-Guided Multi-omics Integration for Drug Response Prediction \\ with Graph Neural Networks

\thanks{\textsuperscript{*} The first two authors contributed equally to this work.}
\thanks{$^{\star}$ J. Cao and J. Wu are the corresponding authors.}
}

\author{
\IEEEauthorblockN{Ruiwei Feng*}
\IEEEauthorblockA{\textit{Coll. of Computer Science and Technology} \\
\textit{Zhejiang University}\\
Hangzhou, China \\
ruiwei\_feng@zju.edu.cn}

\\
\IEEEauthorblockN{Danny Z. Chen, \IEEEmembership{Fellow,~IEEE}}
\IEEEauthorblockA{\textit{Dept. of Computer Science and Engineering} \\
\textit{University of Notre Dame}\\
Notre Dame, USA \\
dchen@nd.edu}

\and

\IEEEauthorblockN{Yufeng Xie*}
\IEEEauthorblockA{\textit{School of Software Technology} \\
\textit{Zhejiang University}\\
Hangzhou, China \\
xieyufeng@zju.edu.cn}
\\
\IEEEauthorblockN{Ji Cao$^{\star}$}
\IEEEauthorblockA{\textit{Coll. of Pharmaceutical Sciences} \\
\textit{Zhejiang University}\\
Hangzhou, China \\
caoji88@zju.edu.cn}

\and

\IEEEauthorblockN{Minshan Lai}
\IEEEauthorblockA{\textit{Polytechnic Institute} \\
\textit{Zhejiang University}\\
Hangzhou, China \\
22060225@zju.edu.cn}

\\
\IEEEauthorblockN{Jian Wu$^{\star}$}
\IEEEauthorblockA{\textit{School of Public Health} \\
\textit{Zhejiang University}\\
Hangzhou, China \\
wujian2000@zju.edu.cn}
}

\maketitle

\begin{abstract}
Accurate drug response prediction (DRP) is a crucial yet challenging task in precision medicine. This paper presents a novel Attention-Guided Multi-omics Integration (AGMI) approach for DRP, which first constructs a Multi-edge Graph (MeG) for each cell line, and then aggregates multi-omics features to predict drug response using a novel structure, called \textbf{G}raph \textbf{e}dge-aware \textbf{Net}work (GeNet). 
For the first time, our AGMI approach explores gene constraint based multi-omics integration for DRP with the whole-genome using GNNs.
Empirical experiments on the CCLE and GDSC datasets show that our AGMI largely outperforms state-of-the-art DRP methods by 8.3\%--34.2\% on four metrics. Our data and code are available at \url{https://github.com/yivan-WYYGDSG/AGMI}.

\end{abstract}


\section{Introduction}
Cancer treatment is a paramount yet challenging task in our society. Precision medicine aims to tailor more effective diagnostic and anti-cancer therapy to each individual patient \cite{porumb2020precision}.
Drug response prediction (DRP) is a critical task in precision medicine, aiming to predict the response of a patient to a given drug. 

\begin{figure*}[h]
    \centering
    \includegraphics[width = 0.85\textwidth]{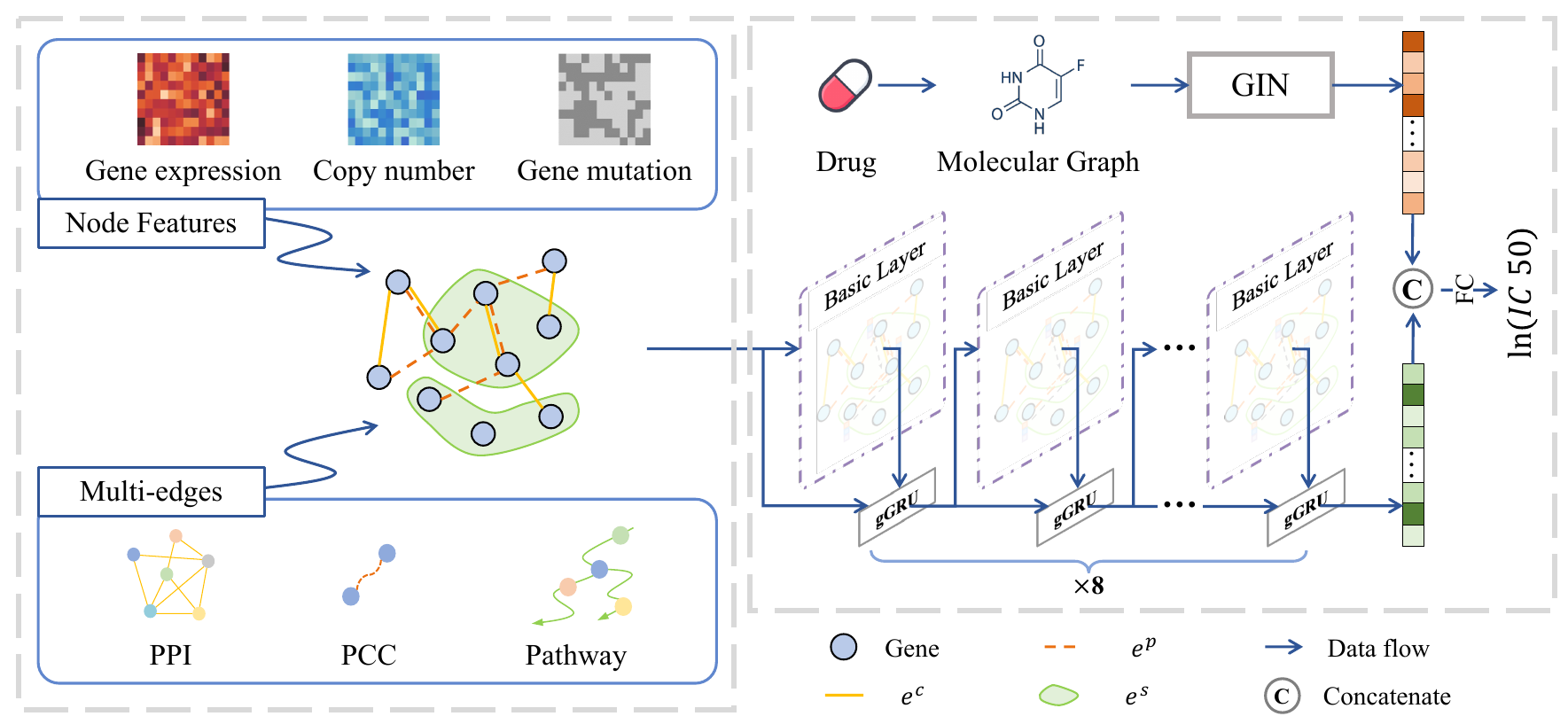}
    \caption{Illustrating our AGMI approach. The left part shows the MeG construction by integrating multi-omics data as node features and multiple types of edges. The right part presents the overall structure of our GeNet.}
    \label{fig_overview}
\end{figure*}

Most known DRP methods take a single type of omics data as input (e.g., gene expression profiles \cite{zhu2020ensemble} or aberrations \cite{chang2018cancer}) to predict drug response. Several methods partially showed that adding more omics data can improve the model predictive power \cite{chen2021survey}. Meanwhile, they revealed some crucial issues in multi-omics integration for DRP.  
Hence, developing an effective multi-omics integration is essential for more accurate drug response prediction. 

To this end, in this paper, we propose a novel Attention-Guided Multi-omics Integration (AGMI) approach for DRP (illustrated in Fig.~\ref{fig_overview}).  Specifically, AGMI first models each cell line as a Multi-edge Graph (MeG), and then conducts attention-guided feature aggregation for multi-omics features with a novel structure, called \textbf{G}raph \textbf{e}dge-aware \textbf{Net}work (GeNet). 
MeG considers the explicit priors of genes, and encodes three basic types of genome features (gene expression, mutation, and CNV) as node features and several other omics data and biological priors on gene relations as multi-edges (e.g., Protein-Protein Interaction (PPI), gene pathways, and Pearson correlation coefficient (PCC) of gene expression).
Then, we develop GeNet based on the Message Passing Neural Networks (MPNNs), by introducing two Gated Recurrent Units (GRUs). One GRU (nGRU) is at the node-level inside the Basic Layer guiding feature extraction of multiple edges, and the other (gGRU) is at the graph-level converging gene features of the whole MeG and fusing multi-scale features. 
Besides, GeNet adopts a Graph Isomorphism Network (GIN) to generate a drug feature vector and concatenate it with a cell line feature vector for final prediction. 

In summary, there are three major contributions in our work:

\begin{itemize}
    \item To our best knowledge, our proposed Attention-Guided Multi-omics Integration (AGMI) approach is the first to explore gene constraint based multi-omics integration for DRP with the whole-genome using GNNs.  
    
    \item Our AGMI approach has two distinct technical aspects. (1) For the first time, AGMI integrates multi-omics data by modeling a cell line as a graph with multiple types of edges (MeG). (2) AGMI adapts MPNNs by introducing a node-level GRU and a graph-level GRU to capture the complex features of MeG, especially being aware of the features of multiple types of edges.
    
    \item Extensive experiments illustrate the superiority of our AGMI approach over state-of-the-art DRP methods. The ablation study verifies the effectiveness of the specific multi-omics integration components of our AGMI. 
\end{itemize}

\section{Approach}

\subsection{Problem Formulation}
In this study, we view the DRP task as a regression problem for the log normalized half-maximal inhibitory concentration ($\ln (\text{IC50})$) values of given cell-drug pairs, following \cite{noghabi2021drug}. For a set of cell lines $\mathcal{N}_c$ and a set of drugs $\mathcal{N}_d$, a drug response matrix $Y \in \mathbb{R} ^ {|\mathcal{N}_c| \times |\mathcal{N}_d|}$ can be defined, where $|\mathcal{N}_c|$ and $|\mathcal{N}_d|$ denote the numbers of cell lines and drugs, respectively. Each entry $y_{i,j}$ in $Y$ denotes the response score $\ln (\text{IC50})$ of cell line $i$ and drug $j$. Thus, the DRP task can be formulated as determining a mapping function $f: \mathcal{N}_c \times \mathcal{N}_d \rightarrow Y$. The main purpose of this study is to explore a function $f$ that can integrate multi-omics data of $\mathcal{N}_c$ in a rigorous and effective way to achieve higher regression accuracy.

\subsection{Attention-Guided Multi-omics Integration (AGMI)}
\label{sec_agmi}
Fig.~\ref{fig_overview} gives an overview of our AGMI, containing two processes: multi-edge graph (MeG) construction and graph aggregation for final prediction.

\subsubsection{Multi-edge Graph (MeG) Construction}
In this work, multi-edge means that there are multiple types of edges between two nodes (genes). Formally, a cell line is modeled by a multi-edge graph $G = (\mathcal{V}, \mathcal{E})$, where $\mathcal{V}$ is the set of nodes and $\mathcal{E}$ is the set of edges.
Each node $v_i \in \mathcal{V}$, $i \in \left\{1, \ldots, N_v \right\}$, represents a gene, with its amount of expression, gene mutation state, and amount of CNV as node features. $N_v$ is the size of $\mathcal{V}$.
An edge $e_{ij} \in \mathcal{E}$ represents a type of relation between two nodes $v_i$ and $v_j$. $N_e$ is the size of $\mathcal{E}$. Specifically, the relations in MeG include protein-protein interactions (proteomics data), gene pathway relations (metabonomics data), and PCC of gene expression (biological prior). Edges built based on these three relations are respectively denoted by  $e^p$, $e^s$, and $e^c$. 
% \textcolor{red}{More details will be presented in our full paper.}

\subsubsection{Graph edge-aware Network (GeNet)}
GeNet aggregates multi-omics data from a cell line graph MeG and extracts drug features from a molecular graph. These two processes are described in Section~\ref{sec_mpnn} and Section~\ref{sec_gin}, respectively. Then, it combines the two feature vectors resulted from the above two processes to predict the final response score with fully connected (FC) layers (described in Section~\ref{sec_fc}). The right part of Fig.~\ref{fig_overview} presents the architecture of GeNet. 

\begin{figure*}[ht]
    \centering
    \includegraphics[width = 0.9\textwidth]{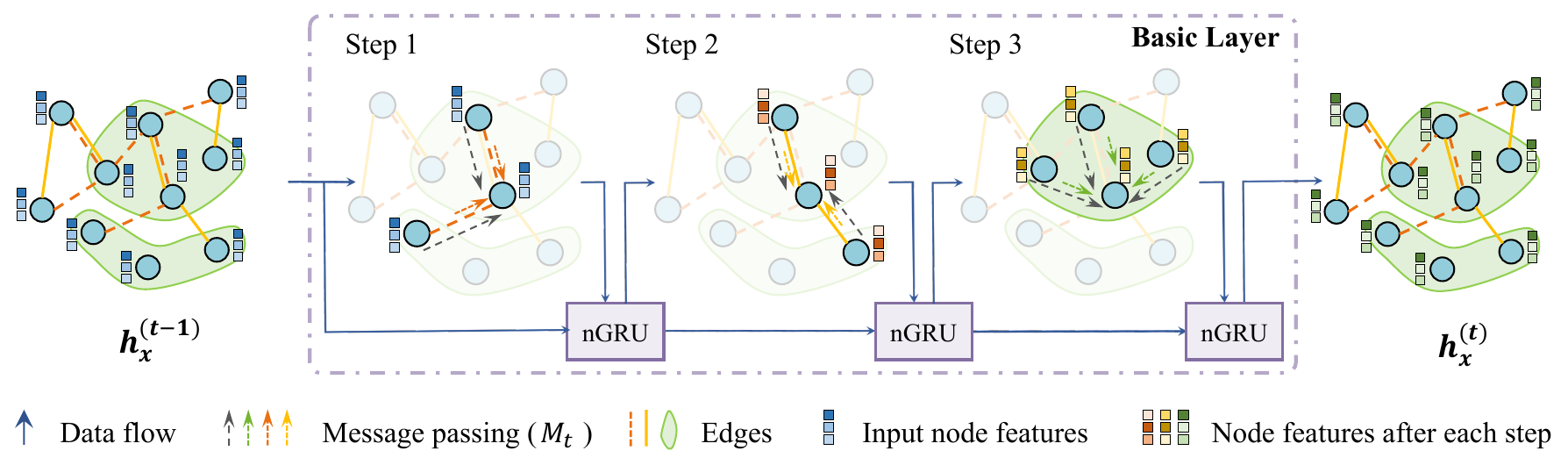}
    \caption{Illustrating the operations in a Basic Layer.}
    \label{fig_basiclayer}
\end{figure*}

\paragraph{Omics-aware Graph Aggregation for Cell Lines}
\label{sec_mpnn}
Following MPNN \cite{gilmer2017neural},
we propose a novel Omics-aware Graph Aggregation in this work. We perform $T = 8$ iterations in the message passing phase, corresponding to $8$ Basic Layers in GeNet.
In each Basic Layer, three steps are conducted to aggregate features for each node from its neighborhood nodes guided by the three types of edges successively (see in Fig.~\ref{fig_basiclayer}).
Besides, we introduce a node-level GRU (nGRU) in the Basic Layer as an attention guidance for aggregating different omics data. Empirically, different Basic Layers aggregate graph features with different receptive fields. To this end, we introduce a graph-level GRU (gGRU) to update vertices of the whole graph by fusing multi-level features, which enables GeNet to maintain features obtained from low-level layers (near the input) to high-level ones.
Finally, we simply use a Neural Network (NN) structure as the readout function, mapping the node features and edge features of the whole graph to a cell line feature vector $z_c \in \mathbb{R}^{128}$. 

\paragraph{A Graph Isomorphism Network (GIN) for Drugs}
\label{sec_gin}
Following previous DRP methods, we collect the SMILES drugs from PubChem 
\cite{kim2016pubchem},
and use the RDKit package to construct molecular graphs \cite{nguyen2021graph}. Atoms are described as nodes, and bonds between any two atoms are described as edges. Due to the state-of-the-art performance of GIN on many molecule-related tasks \cite{nguyen2021graph}, we use a GIN structure to receive molecular graphs and generate the drug representations $z_d \in \mathbb{R}^{128}$.

\paragraph{Fully Connected (FC) Layers for Final Prediction}
\label{sec_fc}
The cell line feature vector $z_c$ and drug representation vector $z_d$ are concatenated for the final prediction, using 3 FC layers. 

\section{Experiments and Results}

\subsection{Datasets}
We conduct experiments using two most commonly used datasets, Cancer Cell Line Encyclopedia (CCLE) \cite{barretina2012cancer} and Genomics of Drug Sensitivity in Cancer (GDSC) \cite{iorio2016landscape}. 
We obtain genomic profiles (including gene expression, gene mutation, and CNV) from the CCLE dataset and response scores of cell-drug pairs (IC50 values) from GDSC. Following \cite{noghabi2021drug}, we apply natural logarithm to the IC50 values and use $ln(IC50)$ values as the regression labels. 
After aligning the genomic profiles and $ln(IC50)$ values for each cell-drug pair, we attain 80758 pieces of records concerning 564 cells and 170 drugs in total. For each cell line, 18499 genes are included after considering missing features. For drugs, only the ones containing PubChem IDs are included. 

As discussed in Section~\ref{sec_agmi}, we include proteomic data (PPI),  metabonomics data (gene pathways), and a biological prior (PCC of gene expression). Specifically, we collect PPI pairs with combined scores from the STRING database \cite{string11} and gene pathways from the GSEA dataset (h, c2, c5, and c6 subsets) \cite{subramanian2005gene}. 
Further, we align multi-omics data of each gene by uniformly using the NCBI ID (version 37) as the uniform identification of genes for different sources.
All the processed data sources of this work are available to researchers interested in multi-omics integration.

\subsection{Experimental Setups}
Following \cite{nguyen2021graph}, we conduct experiments to verify the effectiveness of our approach for known cell lines and drugs. We simply use stratified sampling to split all the cell-drug pairs into training and test sets with a ratio of 9:1. We compute Root Mean Square Error (RMSE), Mean Absolute Error (MAE), Mean Square Error (MSE), and R-squared ($R^2$) to evaluate model performances.

\begin{table}[t]
    \centering
    \caption{Comparison of state-of-the-art DRP methods, including the input omics data and their performances. Various types of omics data are denoted by symbols, as shown in the bottom row. The best and 2nd best results are marked as \textbf{bold} and \underline{underlined}, respectively.}
    \renewcommand\arraystretch{1.2}
    \scalebox{1}{
    \begin{tabular}{p{1.5cm}<{\centering} | p{1.5cm}<{\centering} |cccc}
    \hline
        \multirow{2}*{Method} & \multirow{2}*{Input}  & \multicolumn{4}{c}{Metrics}  \\
        \cline{3-6}
         & &  RMSE & MAE & MSE & $R^2$ \\
         \hline
         DeepDSC & 
            \begin{minipage}[l]{0.1\columnwidth}
        		\centering
        		\raisebox{-.5\height}{\includegraphics[width=0.25\linewidth]{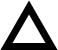}}
        	\end{minipage}
         &  1.0001 & 0.7385 & 1.0005 & 0.8418  \\

         \hline
         CDRScan & 
            \begin{minipage}[l]{0.1\columnwidth}
        		\centering
        		\raisebox{-.5\height}{\includegraphics[width=0.25\linewidth]{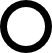}}
        	\end{minipage}
         &  1.0009  & 0.725 &1.0126 & 0.8449  \\
         \hline
         tCNNs & 
            \begin{minipage}[l]{0.1\columnwidth}
        		\centering
        		\raisebox{-.5\height}{\includegraphics[width=0.62\linewidth]{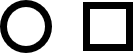}}
        	\end{minipage}
         & 1.0004 & 0.7268 &1.0105 & 0.8448   \\
         \hline
         GraphDRP & 
            \begin{minipage}[l]{0.1\columnwidth}
        		\centering
        		\raisebox{-.5\height}{\includegraphics[width=0.62\linewidth]{figures/symbol/line3.png}}
        	\end{minipage}
         &  \underline{0.9801} & \underline{0.7196} & \underline{0.9595} & 0.8456\\
         \hline
         DeepCDR & 
            \begin{minipage}[l]{0.1\columnwidth}
        		\centering
        		\raisebox{-.5\height}{\includegraphics[width=0.95\linewidth]{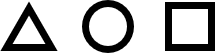}}
        	\end{minipage}
         &  1.0653  & 0.7955 &1.1343 & 0.8199 \\
         \hline
         MOLI &
            \begin{minipage}[l]{0.1\columnwidth}
        		\centering
        		\raisebox{-.5\height}{\includegraphics[width=0.95\linewidth]{figures/symbol/line5.png}}
        	\end{minipage}
         &  1.0178 & 0.7529 & 1.0320 & \underline{0.8478} \\
         \hline
         AGMI (ours) & 
            \begin{minipage}[l]{0.1\columnwidth}
        		\centering
        		\raisebox{-.5\height}{\includegraphics[width=0.29\linewidth]{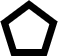}}
        	\end{minipage}
    	& \textbf{0.7943} & \textbf{0.6048} & \textbf{0.6317} & \textbf{0.9184} \\
     \hline
     \multicolumn{6}{c}{
            \begin{minipage}[c]{0.94\columnwidth}
        		\centering
        		\raisebox{-.3\height}{\includegraphics[width=0.98\linewidth]{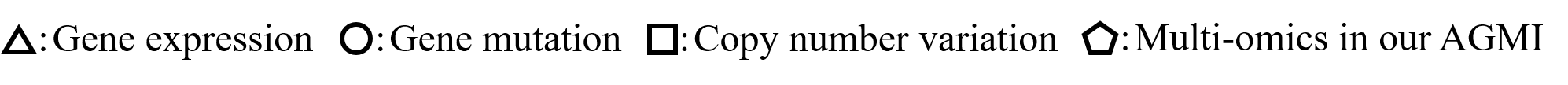}}
        	\end{minipage}
        	}
    \end{tabular}}
    \label{tab_noblind}
\end{table}

\subsection{Results and Analysis}

\subsubsection{Comparison with State-of-the-art DRP Methods}
\label{sec_stfa}
We compare six state-of-the-art DRP methods with our AGMI, including two single-omics methods (DeepDSC and CDRScan) and four multi-omics methods (tCNNs, GraphDRP, DeepCDR, and MOLI). These methods have been well demonstrated in their papers, with specifically selected genes/features and different experimental setups. For fair and convenient comparison, for each method, we feed it with the whole-genome (i.e., 18499 genes), while the type of input omics is the same as in its original paper. Table~\ref{tab_noblind} presents the model input omics and experimental results. Clearly, our AGMI outperforms all these state-of-the-art DRP methods on all the metrics. 

As shown in Table~\ref{tab_noblind}, GraphDRP attains the 2nd best performances in RMSE, PCC, and MSE. Compared to the single-omics methods (i.e., DeepDSC and CDRScan), such superiority may suggest that integrating more types of omics data for DRP is a promising direction. Meanwhile, it is noticeable that
those methods simply integrating multi-omics data as a whole input (i.e., tCNNs and GraphDRP) and concatenating extracted features of multi-omics data (i.e., DeepCDR and MOLI) do not exceed our AGMI. This is possibly due to our two exquisite constructions (MeG and GeNet), which consider explicit priors of genes and potential relations among multi-omics data, and conduct attention-guided multi-omics integration instead of crude concatenation.
Besides, it is worth noting that our AGMI yields higher performances (around 19.0\% on RMSE, 16.0\% on MAE, 34.2\% on MSE, and 8.3\% on $R^2$) than the 2nd best methods. 

\begin{figure}
    \centering
    \includegraphics[width = 0.24\textwidth]{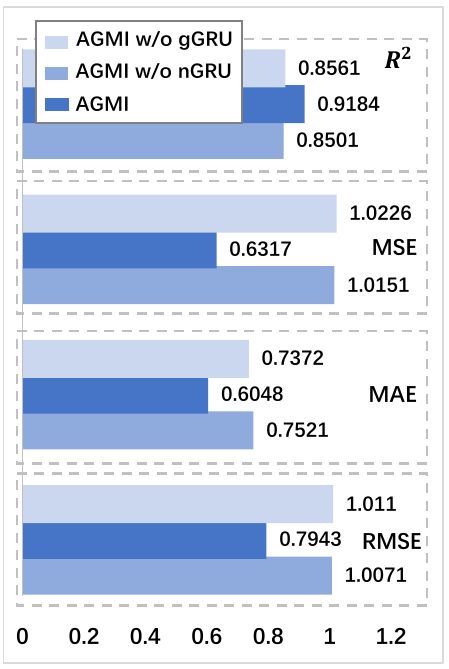}
    \caption{AGMI performances without nGRU or gGRU.}
    \label{fig_abl}
\end{figure}

\subsubsection{Ablation Study}
\label{sec_abl}
This section examines the effectiveness of the two GRUs in our AGMI, by successively removing one of them each time. Fig.~\ref{fig_abl} presents the performances. Note that each GRU module contributes a noticeable improvement on the four metrics.

\subsubsection{Discussion}
Besides its high performances, our AGMI also offers three advantages compared to the known DRP methods:

\begin{itemize}
    \item First, AGMI attempts to utilize the whole-genome information for DRP, instead of empirically selecting several genes as in \cite{sharifi2019moli}. 
    
    \item Second, for the first time, AGMI integrates gene pathways as part of the input information for DRP, instead of only for validation of the computational models. 
    
    \item Third, AGMI is a general framework for all kinds of drugs, instead of specially designed for a specific drug as in \cite{park2021super, kim2021graph}. 
\end{itemize}

\section{Conclusions}
In this paper, we proposed a novel attention-guided multi-omics integration approach, AGMI, for drug response prediction. AGMI models each cell line using a multi-edge graph based on explicit constraints among genes and potential relations among multi-omics data, and aggregates graph features with a sophisticated structure, GeNet. As far as we know, AGMI is the first work to explore multi-omics integration for DRP with the whole-genome using graphs and GNNs. 
Extensive experiments verified that AGMI can achieve much better performances than the known DRP methods, and showed great potential for integrating more types of omics data.
% We believe our work is an important step towards multi-omics integration for DRP task.

\section*{Acknowledgment}
This research was partially supported by the National Research and Development Program of China under grant
2018AAA0102102, the Key R\&D Program of Zhejiang Province No. 2020C03010, the Zhejiang University Education Foundation under grants No. K18-511120-004, No. K17-511120-017, and No. K17-518051-02, the Zhejiang public welfare technology research project under grant No. LGF20F020013, the Leading Innovative and Entrepreneur Team Introduction Program of Zhejiang No.~2019R01007, the Wenzhou Bureau of Science and Technology of China No.~Y2020082, and the Key Laboratory of Medical Neurobiology of Zhejiang Province. D. Z. Chen’s research was supported in part by NSF Grant CCF-1617735. 

\bibliographystyle{ieeetr}
\bibliography{ref}

\end{document}